\documentclass[traditabstract]{aa}

\usepackage{graphicx}
\usepackage{txfonts}
\usepackage{natbib}
\usepackage{color}

\begin{document}

	\title{Mean angular diameters, distances and pulsation modes of the classical Cepheids FF~Aql and T~Vul}
	\titlerunning{CHARA/FLUOR interferometric observations of FF~Aql and T~Vul}

   	\subtitle{CHARA/FLUOR near-infrared interferometric observations}

	\author{ A.~Gallenne\inst{1} \and  
  				P.~Kervella\inst{1} \and
  				A.~M\'erand\inst{2} \and
				H.~McAlister\inst{3}, T.~ten~Brummelaar\inst{3}, V.~Coud\'e du Foresto\inst{1}, J.~Sturmann\inst{3}, L.~Sturmann\inst{3}, N.~Turner\inst{3}, C.~Farrington\inst{3} and P.~J.~Goldfinger\inst{3} 				
  				}
  				
  	\authorrunning{A. Gallenne et al.}

\institute{ LESIA, Observatoire de Paris, CNRS UMR 8109, UPMC, Universit\'e
  Paris Diderot, 5 Place Jules Janssen, F-92195 Meudon, France
  \and European Southern Observatory, Alonso de C\'ordova 3107,
  Casilla 19001, Santiago 19, Chile
  \and Center for High Angular Resolution Astronomy, Georgia State University,
  PO Box 3965, Atlanta, Georgia 30302-3965, USA}
  
  \offprints{A. Gallenne} \mail{alexandre.gallenne@obspm.fr}


 
  \abstract
   {We report the first angular diameter measurements of two classical Cepheids, FF~Aql and T~Vul, that we have obtained with the FLUOR instrument installed at the CHARA interferometric array. We obtain average limb-darkened angular diameters of $\theta_\mathrm{LD} = 0.878 \pm 0.013$\,mas and $\theta_\mathrm{LD} = 0.629 \pm 0.013$\,mas, respectively for FF~Aql and T~Vul. Combining these angular diameters with the HST-FGS trigonometric parallaxes leads to linear radii $R = 33.6 \pm 2.2\,R_\odot$ and $R = 35.6 \pm 4.4\,R_\odot$, respectively. The comparison with empirical and theoretical Period--Radius relations leads to the conclusion that these Cepheids are pulsating in their fundamental mode. The knowledge of the pulsation mode is of prime importance to calibrate the Period--Luminosity relation with a uniform sample of fundamental mode Cepheids.}


 
 \maketitle

%

\section{Introduction}

The direct detection of angular diameter variations of a pulsating star using optical interferometers has been achieved for many stars now \citep[see e.g.][]{Lane_2002_07_0,Kervella_2004_03_0,Merand_2006__0,Davis_2009_04_0,Lacour_2009_12_0}. The combination of angular diameters with radial velocity measurements allows us to accurately estimate their distances in a quasi-geometrical way. This independent distance determination is essential to calibrate the Period-Luminosity (P--L) and Period-Radius (P--R) relations of Cepheids. Direct radius measurements also enable us to compare the theoretical \citep[e.g.][]{Neilson_2010_06_0} and indirect radius estimates \citep[e.g.][]{Groenewegen_2007_11_0}.

Having a direct determination of the diameter, we can also settle whether a Cepheid belongs to the fundamental mode group or not, because its average linear diameter should depart from the classical P--R relation. As different modes yield different relations, it is essential to know the pulsation mode in order to calibrate the P--L relation from a homogeneous sample of Cepheids.

 We present here the first interferometric observations of the Cepheids \object{FF~Aql} and \object{T~Vul}. The  former has a small amplitude (less than 0.5~mag in $V$), a sinusoidal-like light curve, and is therefore generally designated as a s-Cepheid. The s-Cepheids are suspected to be first overtone pulsators \citep[see e.g.][]{Bohm-Vitense_1988_01_0}. T~Vul has the same pulsation period that FF~Aql but is expected to pulsate in the fundamental mode.
 
 These Cepheids also belong to binary systems with a hotter companion. From IUE spectra, \citet{Evans_1990_05_0} detected around FF~Aql a companion with a spectral type between A9~V and F3~V. This corresponds to a magnitude difference with respect to the Cepheid $\Delta K \sim 6.1$--$6.5\,\mathrm{mag}$. \citet{Evans_1992_07_0} found that the companion is a A0.8~V star, also estimated from IUE spectra, and corresponds to a magnitude difference $\Delta K \sim 5.2\,\mathrm{mag}$.

In this paper, we focus on determining the distance and radius of these Cepheids using the interferometric Baade-Wesselink method (IBWM). We also compare the linear radii with those in the literature and with published P--R relations in order to reveal the pulsation mode.

\section{Interferometric observations}

The observations were performed in June 2010 and August 2011 at the CHARA Array \citep{ten-Brummelaar_2005_07_0} in the near infrared $K\arcmin$ band ($1.9 \leqslant \lambda \leqslant 2.3\,\mu\mathrm{m}$) with the Fiber Linked Unit for Optical Recombination \citep[FLUOR;][]{Coude-du-Foresto_2003_02_0,Merand_2006_07_2}. For these observations, we used baselines longer than 200\,m in order to resolved sufficiently well the two targets. The journal of the observations is presented in Table~\ref{table__log}.

\begin{table}[]
\centering
\caption{Calibrated squared visibility measurements of FF~Aql.}
\begin{tabular}{cccccc} 
\hline
\hline
Star		&	MJD	  			&	$\phi$	& 	$B$			&	PA				& 	$V^2 \pm \sigma$		\\
				&						&				&	($m$)		&	($\degr$)	&										\\
\hline
FF~Aql	&	55352.404  	& 0.36  	&  318.98  	&  25.1  		&  $0.2940 \pm 0.0205$  \\ 
		  	&	55352.433  	& 0.36  	&  312.70  	&  19.3  		&  $0.3958 \pm 0.0244$  \\ 
		  	&	55353.431  	& 0.58  	&  273.78  	&  -44.0  	&  $0.4774 \pm 0.0151$  \\ 
		  	&	55355.436  	& 0.03  	&  310.34  	&  16.7  		&  $0.3584 \pm 0.0126$  \\ 
		  	&	55366.374  	& 0.48  	&  218.94  	&  62.9  		&  $0.6192 \pm 0.0175$  \\ 
		  	&	55779.257  	& 0.83  	&  314.15  	&  20.7  		&  $0.3839 \pm 0.0095$  \\ 
		  	&	55781.254  	& 0.27  	&  309.83  	&  72.1  		&  $0.3535 \pm 0.0120$  \\ 
		  	&	55782.317  	& 0.51  	&  230.92  	&   0.1  		&  $0.5813 \pm 0.0159$  \\ 
		  	&	55784.287  	& 0.95  	&  278.37  	&  -48.7  	&  $0.4678 \pm 0.0200$  \\ 
		  	&	55786.266  	& 0.40  	&  308.46  	&  14.4  		&  $0.3506 \pm 0.0136$  \\ 
		  	&	55786.290  	& 0.40  	&  304.97  	&   8.7  		&  $0.3807 \pm 0.0144$  \\ 
		  	&	55790.230  	& 0.28  	&  249.54  	&  76.0  		&  $0.5141 \pm 0.0126$  \\ 
\hline
T~Vul	& 	55779.439 	& 0.25		& 322.81	 & -4.4 & $0.5905\pm0.0227$ \\  
			& 	55779.461 	& 0.25 		& 323.82	 & -9.6 & $0.6059\pm0.0210$ \\  
			& 	55779.487 	& 0.25 		& 325.80	 & -15.7 & $0.6148\pm0.0222$ \\  
			& 	55780.445 	& 0.48 		& 323.12	 & -6.4 & $0.5279\pm0.0294$ \\  
			& 	55780.496 	& 0.48 		& 326.81	 & 18.2 & $0.5335\pm0.0257$ \\  
			& 	55781.403 	& 0.69 		& 283.61	 & 55.9 & $0.6252\pm0.0172$ \\  
			& 	55781.423 	& 0.69 		& 270.23	 & 50.3 & $0.7054\pm0.0213$ \\  
			& 	55782.479 	& 0.93 		& 246.61	 & -18.3 & $0.7524\pm0.0211$ \\  
			& 	55784.382 	& 0.36 		& 274.57	 & -51.5 & $0.6772\pm0.0332$ \\  
			& 	55787.478 	& 0.06 		& 326.91	 & -18.5 & $0.6241\pm0.0256$ \\  
\hline
\end{tabular}
\tablefoot{MJD is the date of the observations (Modified Julian Date), $\phi$ the phase of pulsation, $B$ the telescope projected separation, PA the baseline projection angle, and $V^2$ the squared visibility.}
\label{table__log}
\end{table}

\begin{table}[]
\centering
\caption{Calibrators used for the observations.}
\begin{tabular}{cccccc} 
\hline
\hline
Calibrator 						&	$m_\mathrm{V}$	& 	$m_\mathrm{K}$	&	Spec. Type	& 	$\theta_\mathrm{UD}$	&	$\gamma$		\\
		  											&								&								&						&	(mas)							&	($\degr$)	\\
\hline		
\multicolumn{6}{c}{FF~Aql} \\
\object{HD~172169}  	& 	6.7	  					&  4.1						&  K4III  			&  $0.942 \pm 0.012$ 		&	13.1			\\ 
\object{HD~169113}		&	7.1						&	3.9						&	K1III				&	$0.816 \pm 0.011$		&	13.3			\\
\object{HD~187193}		&	6.0						&	3.9						&	K0II-III			&	$0.893 \pm 0.012$		&	14.0			\\
\hline
\multicolumn{6}{c}{T~Vul} \\
\object{HD~198330}  	& 	7.3	  					&  3.8						&  K4III  			&  $0.913 \pm 0.012$ 		&	2.6			\\ 
\object{HD~201051}		&	6.1						&	4.0						&	K0II-III			&	$0.827 \pm 0.011$		&	3.6			\\
\object{HD~192944}		&	5.3						&	3.0						&	G8III				&	$1.085 \pm 0.015$		&	8.5			\\
HD~187193					&	6.0						&	3.9						&	K0II-III			&	$0.893 \pm 0.012$		&	14.5			\\
\hline
\end{tabular}
\tablefoot{$m_\mathrm{V}, m_\mathrm{K}$: magnitudes in $V$ and $K$ bands; $\theta_\mathrm{UD}$: uniform disk angular diameter in $K$ band; $\gamma$: angular distance to the Cepheid.}
\label{table__calibrator}
\end{table}

The squared visibility of the fringes has been estimated from the raw data using the FLUOR data reduction software \citep{Coude-du-Foresto_1997_02_0,Merand_2006_07_2}, based on the integration of the fringes power spectrum. The raw squared visibilities have then been calibrated using resolved calibrator stars, chosen from the catalogue of \citet{Merand_2005_04_0}, using criteria defined to minimize the calibration bias and maximize the signal-to-noise ratio. The calibrators used for these observations are listed in Table~\ref{table__calibrator}. They were observed immediately before and/or after the Cepheid in order to monitor the interferometric transfer function of the instrument. The error introduced by the uncertainty on each calibrator's estimated angular diameter has been propagated using the formalism developed by \citep{Perrin_2003_03_0}. The final uncertainty of the calibrated squared visibility includes statistical (from the dispersion of the raw squared visibilities obtained during the observations) and systematic errors (from the error bars on the calibrators).

For each night we observed a Cepheid, we determined a uniform disk diameter based on one or several squared visibility measurements (Table~\ref{table__log}). Each night was assigned a unique pulsation phase established using the average date of observation and the \citet{Samus_2009_01_0} ephemeris: $P = 4.470916$ days and $T_\mathrm{0} = 2441576.4280$ for FF~Aql, and $P = 4.435462$ days and $T_\mathrm{0} = 2441705.1210$ for T~Vul. 

\section{The interferometric Baade--Wesselink method}

The IBWM makes use of radial velocity and angular diameter measurements of the star during its pulsation cycle. This involves a good phase coverage and the highest possible angular resolution. The integral over time of the radial velocity is then directly linked to the angular diameter variation.

The presence of a companion can falsify the results if its magnitude is close to the one of the Cepheid, and if the orbital effect is detectable in the spectroscopic measurements. In our case, these main sequence companions are faint compared to the Cepheids, and do not affect the photometric signals. Only the radial velocities are affected by the presence of the companions and have to be corrected.

\subsection{Radial velocity integrations}

Among published radial velocities data of FF~Aql, we chose \citet{Evans_1990_05_0} for two reasons. First, the orbital motion causes by the presence of the companion was already removed from measurements, only leaving the displacement due to the pulsating photosphere.  Secondly, the radial velocities were extracted using the cross-correlation method. As shown by \citet{Nardetto_2004_12_0}, the method used can affect the distance determination via the choice of the projection factor (hereafter $p$-factor), which we will introduce in Sec~\ref{subsection__distance_determination}. For T~Vul, we use the radial velocities of \citet{Bersier_1994_11_0}, also from the cross-correlation method. The amplitude of the orbital motion was not detected with this data so we did not apply any corrections due to the presence of the companion. For both Cepheids, we redetermined the pulsation phases using the ephemeris presented previously.

\begin{figure*}[!ht]
\centering
\resizebox{\hsize}{!}{\includegraphics{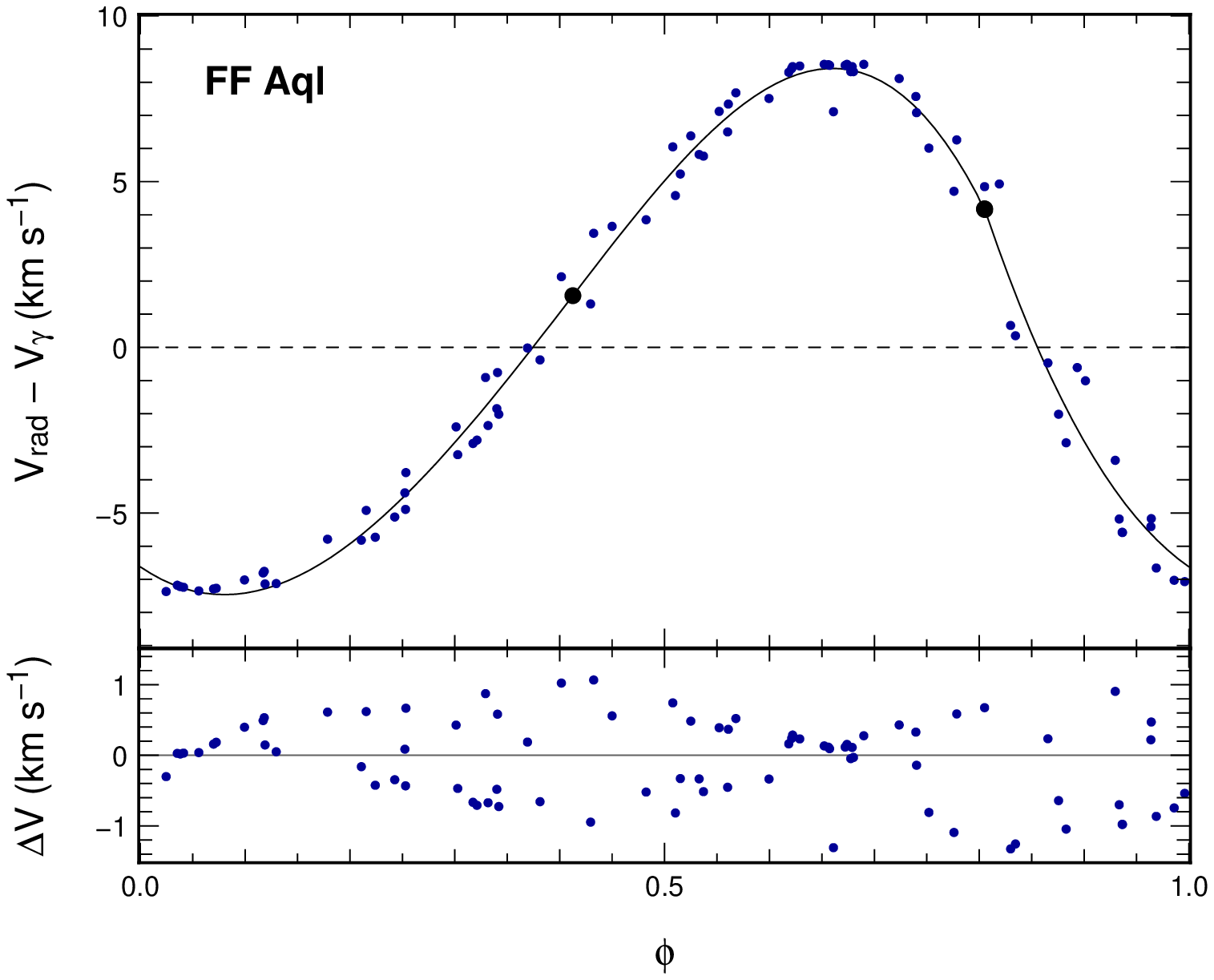}\hspace{.5cm}
\includegraphics{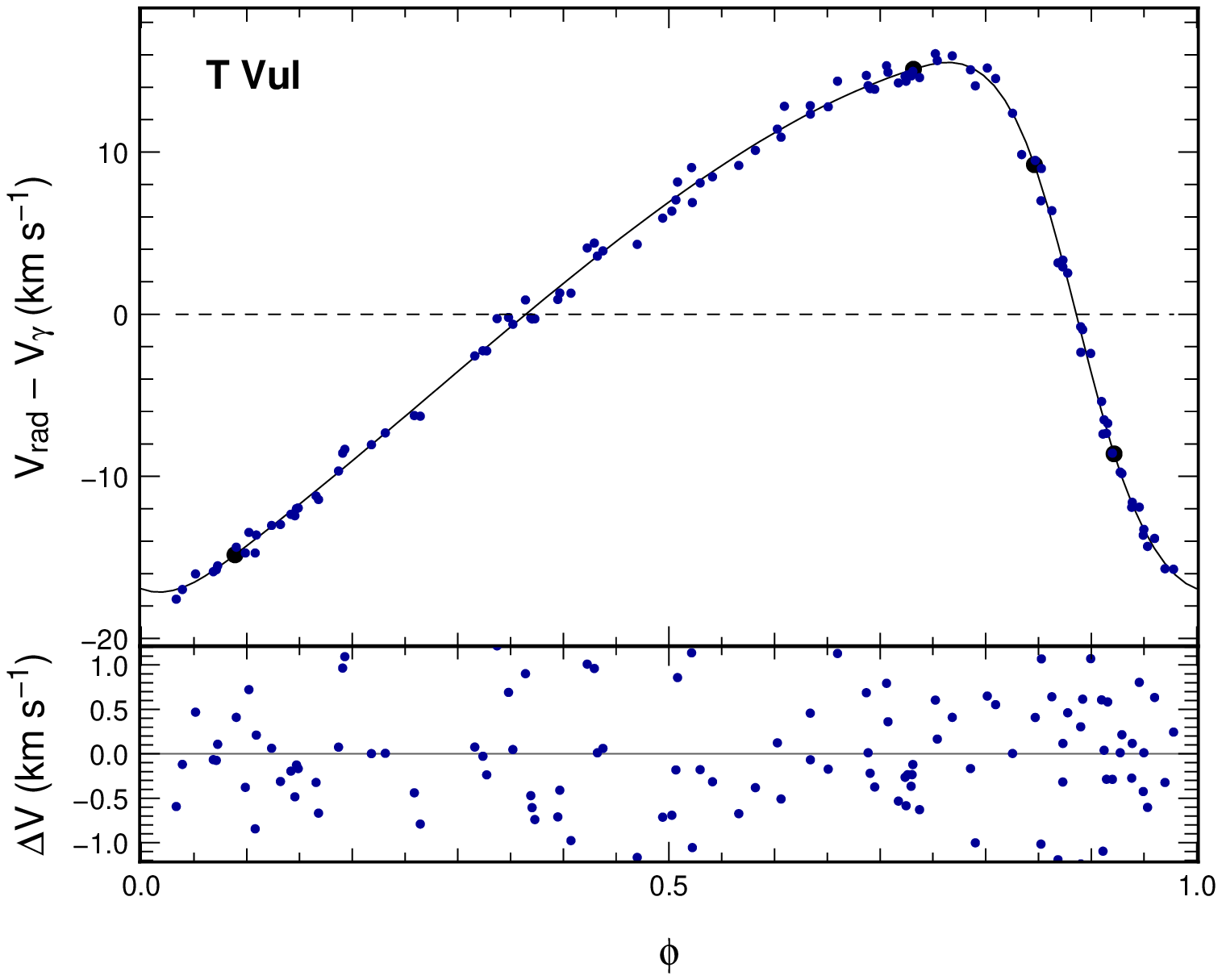}}
\caption{Radial velocity of FF~Aql and T~Vul. The solid line is the periodic spline function defined with four adjustable floating nodes (large black dots). The lower panel displays the residuals of the fit.}
\label{image__rv}
\end{figure*}

The radial velocities, which were acquired at irregular phases, must be numerically integrated to get the radial displacement. The best way to achieve a robust integration and avoid noisy data is to interpolate the data before integration. For this purpose, the radial velocities were smoothly interpolated using a periodic cubic spline function, defined by floating nodes. This method was already used by \citet{Merand_2007_08_0} and we refer the reader to this paper for a detailed explanation. The interpolated curves for our two Cepheids are presented in Fig.~\ref{image__rv}. The dispersion of the residuals are $\sigma = 61\,\mathrm{m\,s^{-1}}$ and $\sigma = 68\,\mathrm{m\,s^{-1}}$, respectively for T~Vul and FF~Aql.

\begin{figure*}[!ht]
\centering
\resizebox{\hsize}{!}{\includegraphics{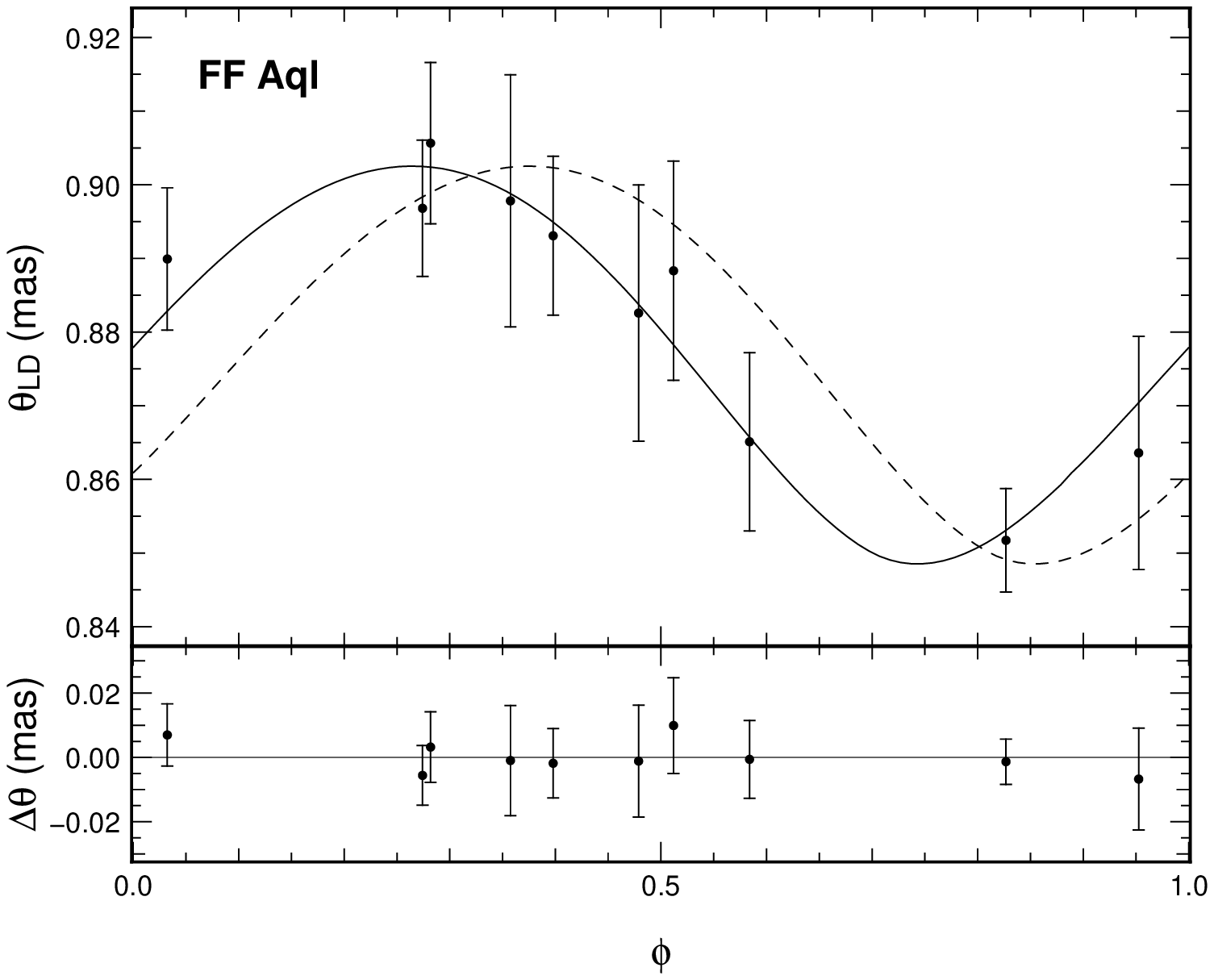}\hspace{.5cm}
\includegraphics{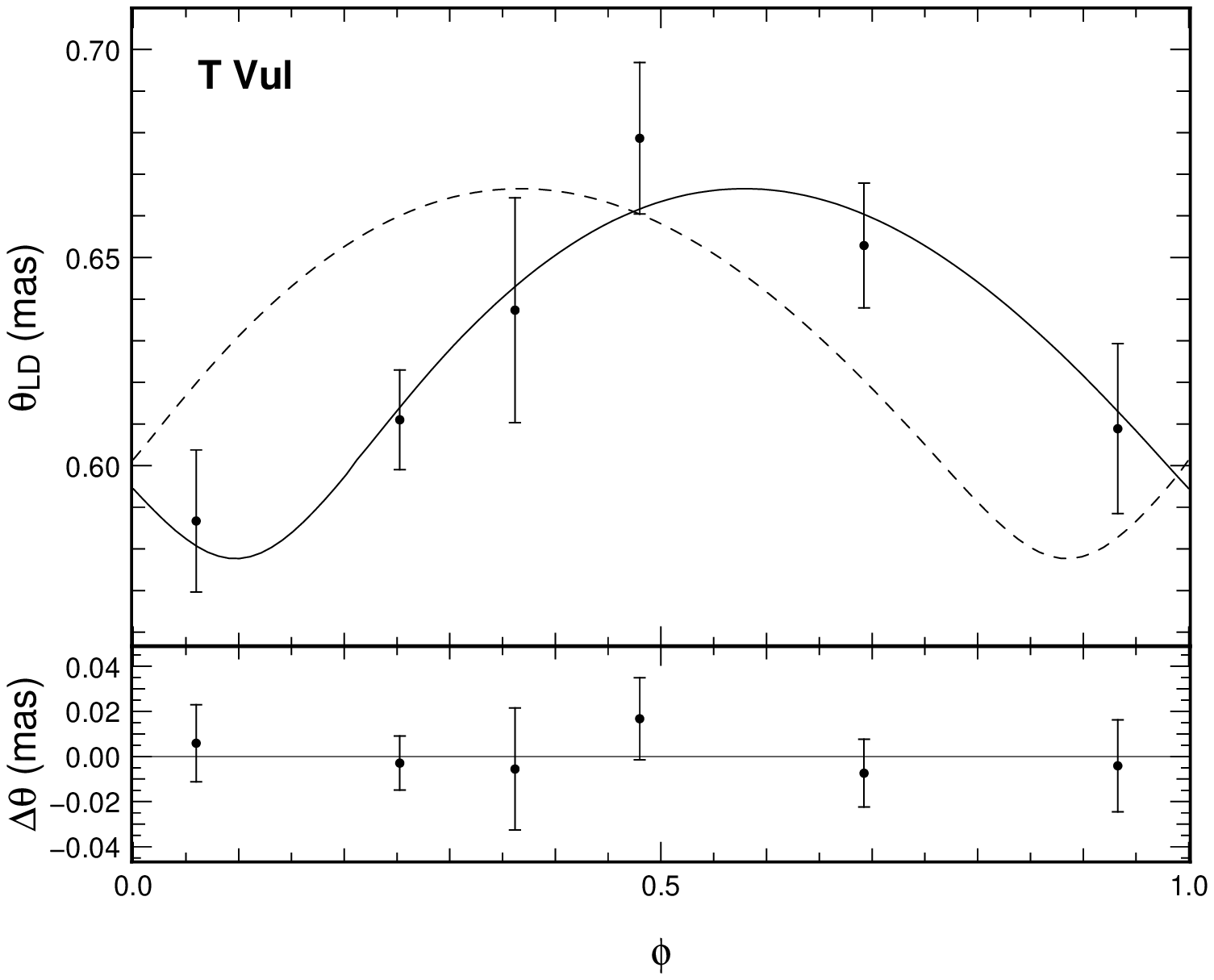}}
\caption{Uniform disk angular diameter variation of FF~Aql and T~Vul. The solid line is the integration of the radial velocity curve with the distance, average angular diameter, and period change as fitted parameters. The dashed line includes the period change. The lower panel displays the residuals of the fit.}
\label{image__diam}
\end{figure*}

\subsection{Distance determination}
\label{subsection__distance_determination}

Our angular diameter measurements were then fitted to the radial displacements to get the distance $d$:
\begin{equation}
\label{equation__BW}
\theta_\mathrm{UD}(T) - \theta_\mathrm{UD}(0) = -2 \frac{kp}{d} \int_0^T (v_\mathrm{r}(t) - v_\gamma) dt,
\end{equation}
where $p$ is the so-called $p$-factor, defined as the ratio between the pulsation velocity and the radial velocity (measured by spectroscopy), $\theta_\mathrm{UD}$ the interferometric angular diameter for a uniform disk model, $k$ the ratio between the uniform and the limb-darkened stellar diameter, and $v_\gamma$ the systemic velocity.

The parameters of the fit are the average angular diameter $\theta_\mathrm{UD}(0)$, the quantity $d/kp$ and a possible variation in the pulsation period between interferometric and spectroscopic observations.

The $p$-factor depends on detailed modelling of stellar atmosphere, but such models are only available for a few Cepheids and we conventionally choose a multiplicative constant factor. \citet{Nardetto_2004_12_0} showed that the choice of a constant factor instead of a time-dependent ones gives a systematic error in the final distance of $\sim 0.2$\,\%, which is below the best relative precision of current distance estimates. \citet{Merand_2005_07_0} presented the unique measured value $p = 1.27 \pm 0.06$ for $\delta$~Cep. Although there is no agreement in the literature on the optimum value of the $p$-factor, it is clear that it depends on each Cepheid. Different authors use either constant p-factor values (ranging from 1.27 to 1.5), or a dependence with the pulsation period. The period dependence is likely to reflect the dependence with the limb darkening. For this work, we chose the latest $p$-factor relation $p = 1.550_\mathrm{\pm 0.04} - 0.186_\mathrm{\pm 0.06} \log P$ of \citet{Storm_2011_10_0}, because this is an empirical relation based on the largest sample of Cepheids, and covering the widest range of periods. We have for both stars $p = 1.43 \pm 0.06$.

The $k$ parameter is generally determined assuming a limb-darkened disk model and is expected to vary with the pulsation phase. However, in the near-IR, \citet{Marengo_2003_06_0} showed from hydrodynamic models in spherical geometry that $k$ is constant with the pulsation at a level of 0.2\,\%. From hydrostatic models, \citet{Merand_2008__0} found a variation of $k$ of the order of 0.4\,\%. Our interferometric measurements are not sensitive to this variation and we therefore assumed a constant limb-darkening coefficient. The hydrostatic model computed from Claret coefficients \citep{Claret_2000_11_0}, with the average stellar parameters $T_\mathrm{eff} =  6000\,\mathrm{K}, \log g = 2$, Fe/H = 0.0 and $V_\mathrm{t} = 4\,\mathrm{km\,s}^{-1}$ from \citet{Luck_2008_07_0} for FF~Aql and \citet{Andrievsky_2005_10_0} for T~Vul, corresponds to $k = 0.985$. The IBWM fit yields for FF~Aql a distance $d = 363 \pm 73 \,$pc and an average angular uniform disk diameter $\theta_\mathrm{UD} = 0.865 \pm 0.013$\,mas, with a total reduced $\chi^2$ of 0.4. For T~Vul, we obtain $d = 415 \pm 100 \,$pc and $\theta_\mathrm{UD} = 0.620 \pm 0.012$\,mas, with a total reduced $\chi^2$ of 2.5. These results are summarized in Table~\ref{table__bw_results} and the angular diameter variations are plotted in Fig.~\ref{image__diam}. We also allowed a change in the period for FF~Aql and T~Vul of $\sim 14$\,s and $\sim -26$\,s, respectively.

\begin{table*}[!ht]
\centering
\caption{Cepheid average angular diameters and distances determined through the application of the interferometric BW method.}
\begin{tabular}{ccccccc} 
\hline
\hline
Star		&	$\overline{\theta_\mathrm{UD}}$	&	$\overline{\theta_\mathrm{LD}}$	& 	$d$	& $d_\mathrm{HST}$	&	$R$	&	$\chi_\mathrm{r}^2$	\\
				&			(mas)									&		(mas)										&	(pc)	& (pc)	&	($R_\odot$)	&						\\
\hline
FF~Aql	&	$0.865 \pm 0.013$	   &	$0.878 \pm 0.013$	&	$363 \pm 73$	& $356 \pm 23$	&	$33.6 \pm 2.2$	&	0.4	\\ 
T~Vul	&	$0.620 \pm 0.012$		&	$0.629 \pm 0.013$	&	$415 \pm 100$	&	$526 \pm 64$	&	$35.6 \pm 4.4$	&	2.5	\\
\hline
\end{tabular}
\tablefoot{The linear radius $R$ were estimated using the more accurate distances $d_\mathrm{HST}$ derived by \citet{Benedict_2007_04_0}.}
\label{table__bw_results}
\end{table*}

\subsection{Linear radius}

We transformed the uniform disk diameters to limb-darkened disk diameters using the value of $k$ previously chosen, we get $\theta_\mathrm{LD} = 0.878 \pm 0.013$\,mas for FF~Aql and $\theta_\mathrm{LD} = 0.629 \pm 0.013$\,mas for T~Vul. These measured values, which have an accuracy $\sim1.5$\,\%, are in good agreement with those predicted from IR surface brightness method 
\citep[e.g.][]{Moskalik_2005_06_0,Groenewegen_2007_11_0}. The fitted distance of FF~Aql is also consistent with the derived value $d = 356 \pm 23$\,pc from the HST-FGS parallax of \citet{Benedict_2007_04_0}. The same authors give for T~Vul $d = 526 \pm 64$\,pc, that is $\sim 27$\,\% larger than our value, but within $1.1\sigma$. As the values from \citet{Benedict_2007_04_0} are more accurate and do not depend on a $p$-factor assumption (compared to our fitted distances), we preferred them to evaluate the linear radius. We therefore used the HST-FGS distances in the BW fit and obtained $R = 33.6 \pm 2.2\,R_\odot$ and $R = 35.6 \pm 4.4\,R_\odot$, respectively for FF~Aql and T~Vul. 

In our analysis, we did not take into account a possible circumstellar envelope (CSE) emission, that could lead to an overestimate of the angular diameter. From the spectral energy distribution of FF~Aql given by \citet{Gallenne_2011_11_0}, the infrared excess caused by the CSE appears around $10\,\mu\mathrm{m}$, while it is negligible at $2.2\,\mu\mathrm{m}$. We assumed this is also the case for T~Vul.

\subsection{$p$-factor}

We can use the HST-FGS distances in Equ.~\ref{equation__BW} to directly fit the $p$-factor, as already done by \citet{Merand_2005_07_0} for $\delta$~Cep. We get $p = 1.40 \pm 0.28$ and $p = 1.81 \pm 0.44$, for FF~Aql and T~Vul respectively. These estimates can be compared with the empirical and theoretical $p$--P relations, respectively from \citet{Storm_2011_10_0} and \citet{Nardetto_2009_05_0}. For FF~Aql, the $p$-factor is compatible with the value derived from the empirical relation, while the estimate of T~Vul is within $1\sigma$. Although at odds with the theoretical relation, our estimates are not accurate enough to constrain it. 

A direct determination of the $p$-factor is a powerful constraint to numerical atmosphere models of Cepheids, however our work does not contain enough measurements to have an accurate estimate of this parameter for these stars.

\section{Period--Radius relation}

The estimated linear radii can be compared with empirical and theoretical Period--Radius (P--R) relations, that are of the form $\log R = a\,\log P + b$. We chose the recent theoretical calculations from \citet{Neilson_2010_06_0}, who used a pulsation-driven mass loss model. We also selected the recent works of \citet{Groenewegen_2007_11_0}, who used Cepheids with known distances and angular diameters to derive an empirical P--R relation. These two relations are only valid for fundamental mode Cepheids.

\begin{figure}[!ht]
\centering
\resizebox{\hsize}{!}{\includegraphics{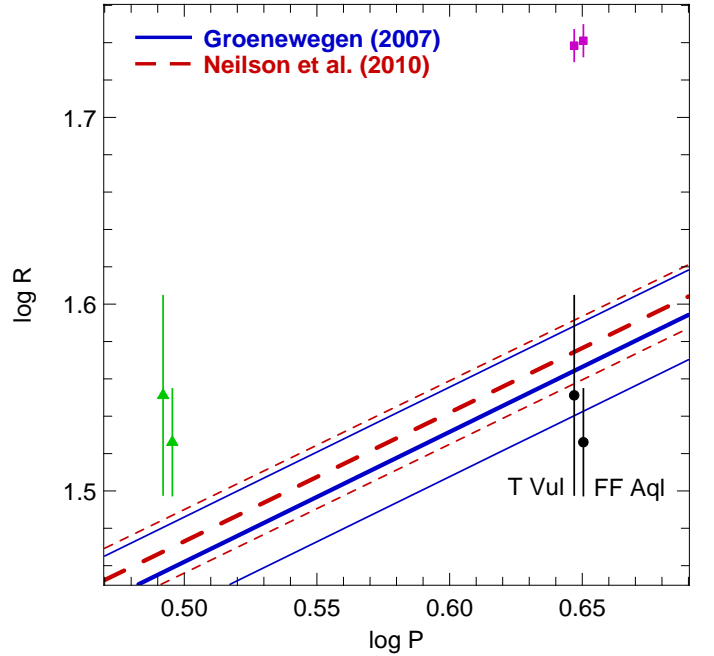}}
\caption{Period--Radius relation for fundamental mode Cepheids. The colour curves represent empirical or theoretical P--R relations from the literature. Each P--R intrinsic dispersion is plotted with a thinner curve, with the same colour and symbol. The dots are our estimated linear radii, the triangles are converted for first overtone pulsation, and the squares represent the theoretical first overtone radii.}
\label{image__PR}
\end{figure}

In principle, the position of a Cepheid in the P--R diagram can reveal its pulsation mode, if the estimated radius is sufficiently accurate. We plotted in Fig.~\ref{image__PR} the P--R relations (thick curves) and our estimated linear radii. The intrinsic dispersion of each relation is represented with thinner curves (with the same colour and symbol). We notice that both Cepheids are consistent (within the intrinsic dispersion) with stars pulsating in a fundamental mode. We also plotted with triangles the radii assuming an overtone pulsation, using the ratio of the overtone period $P_\mathrm{1}$ to the fundamental period $P_\mathrm{0}$ given by $P_\mathrm{1}/P_\mathrm{0} = 0.720 - 0.027\,\log P_\mathrm{0}$ \citep{Alcock_1995_04_0}. In addition, we represented with squares the radii of the star assuming first overtone radii, by using the theoretical first overtone P--R relation of \citet{Bono_2001_08_0}. The overtone mode seems not compatible with the P--R relations, and therefore suggest that FF~Aql and T~Vul are pulsating in the fundamental mode.

\section{Conclusions and discussion}

We report the first interferometric observations of two classical Cepheids, FF~Aql and T~Vul, with the CHARA/FLUOR instrument. The angular diameter variations were monitored over the pulsation cycle and combined with previously published radial velocity measurements to get an independent estimate of the distance and mean angular diameter. The precision achieved is $\sim 2$\,\% on their average angular diameter estimates. 

We found for FF~Aql a good agreement ($0.3\sigma$) between our fitted distance and the HST-FGS estimate \citep{Benedict_2007_04_0}. For T~Vul, a discrepancy is found between the two estimates, our value being $\sim 27$\,\% smaller, but within $1.1\sigma$. The main source of bias in the use of the BW method is the choice of the $p$-factor, and that could explain the discrepancy encountered. We can also note that more interferometric measurements, with higher accuracy, would also reduce this discrepancy. To avoid a possible bias on the distance due to the $p$-factor assumption, we coupled our angular diameters with existing trigonometric parallaxes to evaluate the linear radii. 

When we compared the linear radii with empirical and theoretical P--R relations, we showed that these two short period Cepheids are pulsating in their fundamental mode.  Our work confirms the pulsation mode of T~Vul, already flagged as fundamental pulsator by several authors \citep[e.g.][ and references therein]{Groenewegen_2000_11_0}. However this is not the case for FF~Aql. This $s$-Cepheid, with small light amplitude and practically sinusoidal light curve, is usually considered as pulsating in the first overtone \citep[e.g.][]{Sachkov_1997_10_0}. As the pulsation mode is generally identified by Fourier decomposition of the light curve or the radial velocity curve, the presence of a companion around FF~Aql could lead to an error in the determination of the pulsation mode. If the difference between the magnitude of the Cepheid and that of its companion is small, the apparent magnitude of the Cepheid will be falsified. The radial velocity measurements can also be altered because of the orbital effect.

The knowledge of the pulsation mode is of prime importance to calibrate the P--L relation with an uniform sample of fundamental mode Cepheids, since different modes will yield different relations.


\begin{acknowledgements}
The authors would like to thank the CHARA Array and Mount Wilson Observatory staff for their support. Research conducted at the CHARA Array is funded by the National Science Foundation through NSF grant AST-0908253, by Georgia State University, the W. M. Keck Foundation, the Packard Foundation, and the NASA Exoplanet Science Institute. We received the support of PHASE, the high angular resolution partnership between ONERA, Observatoire de Paris, CNRS, and University Denis Diderot Paris 7. This work made use of the SIMBAD and VIZIER astrophysical database from CDS, Strasbourg, France and the bibliographic informations from the NASA Astrophysics Data System. Data processing for this work have been done using the Yorick language which is freely available at http://yorick.sourceforge.net/.
\end{acknowledgements}


\bibliographystyle{aa}   
\bibliography{../../../bibliographie}

\begin{thebibliography}{35}
\expandafter\ifx\csname natexlab\endcsname\relax\def\natexlab#1{#1}\fi

\bibitem[{{Alcock} {et~al.}(1995){Alcock}, {Allsman}, {Axelrod}, {Bennett},
  {Cook}, {Freeman}, {Griest}, {Marshall}, {Peterson}, {Pratt}, {Quinn},
  {Reimann}, {Rodgers}, {Stubbs}, {Sutherland}, \& {Welch}}]{Alcock_1995_04_0}
{Alcock}, C., {Allsman}, R.~A., {Axelrod}, T.~S., {et~al.} 1995, \aj, 109, 1653

\bibitem[{{Andrievsky} {et~al.}(2005){Andrievsky}, {Luck}, \&
  {Kovtyukh}}]{Andrievsky_2005_10_0}
{Andrievsky}, S.~M., {Luck}, R.~E., \& {Kovtyukh}, V.~V. 2005, \aj, 130, 1880

\bibitem[{{Benedict} {et~al.}(2007){Benedict}, {McArthur}, {Feast}, {Barnes},
  {Harrison}, {Patterson}, {Menzies}, {Bean}, \&
  {Freedman}}]{Benedict_2007_04_0}
{Benedict}, G.~F., {McArthur}, B.~E., {Feast}, M.~W., {et~al.} 2007, \aj, 133,
  1810

\bibitem[{{Bersier} {et~al.}(1994){Bersier}, {Burki}, {Mayor}, \&
  {Duquennoy}}]{Bersier_1994_11_0}
{Bersier}, D., {Burki}, G., {Mayor}, M., \& {Duquennoy}, A. 1994, \aaps, 108,
  25

\bibitem[{{Bohm-Vitense}(1988)}]{Bohm-Vitense_1988_01_0}
{Bohm-Vitense}, E. 1988, \apjl, 324, L27

\bibitem[{{Bono} {et~al.}(2001){Bono}, {Caputo}, \& {Marconi}}]{Bono_2001_08_0}
{Bono}, G., {Caputo}, F., \& {Marconi}, M. 2001, \mnras, 325, 1353

\bibitem[{{Claret}(2000)}]{Claret_2000_11_0}
{Claret}, A. 2000, \aap, 363, 1081

\bibitem[{{Coud{\'e} du Foresto} {et~al.}(2003){Coud{\'e} du Foresto}, {Borde},
  {Merand}, {Baudouin}, {Remond}, {Perrin}, {Ridgway}, {ten Brummelaar}, \&
  {McAlister}}]{Coude-du-Foresto_2003_02_0}
{Coud{\'e} du Foresto}, V., {Borde}, P.~J., {Merand}, A., {et~al.} 2003, in
  SPIE Conference Series, ed. {W.~A.~Traub}, Vol. 4838, 280--285

\bibitem[{{Coud{\'e} du Foresto} {et~al.}(1997){Coud{\'e} du Foresto},
  {Ridgway}, \& {Mariotti}}]{Coude-du-Foresto_1997_02_0}
{Coud{\'e} du Foresto}, V., {Ridgway}, S., \& {Mariotti}, J. 1997, \aaps, 121,
  379

\bibitem[{{Davis} {et~al.}(2009){Davis}, {Jacob}, {Robertson}, {Ireland},
  {North}, {Tango}, \& {Tuthill}}]{Davis_2009_04_0}
{Davis}, J., {Jacob}, A.~P., {Robertson}, J.~G., {et~al.} 2009, \mnras, 394,
  1620

\bibitem[{{Evans}(1992)}]{Evans_1992_07_0}
{Evans}, N.~R. 1992, \aj, 104, 216

\bibitem[{{Evans} {et~al.}(1990){Evans}, {Welch}, {Scarfe}, \&
  {Teays}}]{Evans_1990_05_0}
{Evans}, N.~R., {Welch}, D.~L., {Scarfe}, C.~D., \& {Teays}, T.~J. 1990, \aj,
  99, 1598

\bibitem[{{Gallenne} {et~al.}(2011){Gallenne}, {Kervella}, \&
  {M{\'e}rand}}]{Gallenne_2011_11_0}
{Gallenne}, A., {Kervella}, P., \& {M{\'e}rand}, A. 2011, \aap, 538, A24

\bibitem[{{Groenewegen}(2000)}]{Groenewegen_2000_11_0}
{Groenewegen}, M.~A.~T. 2000, \aap, 363, 901

\bibitem[{{Groenewegen}(2007)}]{Groenewegen_2007_11_0}
{Groenewegen}, M.~A.~T. 2007, \aap, 474, 975

\bibitem[{{Kervella} {et~al.}(2004){Kervella}, {Nardetto}, {Bersier},
  {Mourard}, \& {Coud{\'e} du Foresto}}]{Kervella_2004_03_0}
{Kervella}, P., {Nardetto}, N., {Bersier}, D., {Mourard}, D., \& {Coud{\'e} du
  Foresto}, V. 2004, \aap, 416, 941

\bibitem[{{Lacour} {et~al.}(2009){Lacour}, {Thi{\'e}baut}, {Perrin}, {Meimon},
  {Haubois}, {Pedretti}, {Ridgway}, {Monnier}, {Berger}, {Schuller},
  {Woodruff}, {Poncelet}, {Le Coroller}, {Millan-Gabet}, {Lacasse}, \&
  {Traub}}]{Lacour_2009_12_0}
{Lacour}, S., {Thi{\'e}baut}, E., {Perrin}, G., {et~al.} 2009, \apj, 707, 632

\bibitem[{{Lane} {et~al.}(2002){Lane}, {Creech-Eakman}, \&
  {Nordgren}}]{Lane_2002_07_0}
{Lane}, B.~F., {Creech-Eakman}, M.~J., \& {Nordgren}, T.~E. 2002, \apj, 573,
  330

\bibitem[{{Luck} {et~al.}(2008){Luck}, {Andrievsky}, {Fokin}, \&
  {Kovtyukh}}]{Luck_2008_07_0}
{Luck}, R.~E., {Andrievsky}, S.~M., {Fokin}, A., \& {Kovtyukh}, V.~V. 2008,
  \aj, 136, 98

\bibitem[{{Marengo} {et~al.}(2003){Marengo}, {Karovska}, {Sasselov},
  {Papaliolios}, {Armstrong}, \& {Nordgren}}]{Marengo_2003_06_0}
{Marengo}, M., {Karovska}, M., {Sasselov}, D.~D., {et~al.} 2003, \apj, 589, 968

\bibitem[{{M{\'e}rand}(2008)}]{Merand_2008__0}
{M{\'e}rand}, A. 2008, in EAS Publications Series, Vol.~28, EAS Publications
  Series, ed. S.~{Wolf}, F.~{Allard}, \& P.~{Stee}, 53--59

\bibitem[{{M{\'e}rand} {et~al.}(2007){M{\'e}rand}, {Aufdenberg}, {Kervella},
  {Foresto}, {ten Brummelaar}, {McAlister}, {Sturmann}, {Sturmann}, \&
  {Turner}}]{Merand_2007_08_0}
{M{\'e}rand}, A., {Aufdenberg}, J.~P., {Kervella}, P., {et~al.} 2007, \apj,
  664, 1093

\bibitem[{{M{\'e}rand} {et~al.}(2005{\natexlab{a}}){M{\'e}rand}, {Bord{\'e}},
  \& {Coud{\'e} Du Foresto}}]{Merand_2005_04_0}
{M{\'e}rand}, A., {Bord{\'e}}, P., \& {Coud{\'e} Du Foresto}, V.
  2005{\natexlab{a}}, \aap, 433, 1155

\bibitem[{{M{\'e}rand} {et~al.}(2006{\natexlab{a}}){M{\'e}rand}, {Coud{\'e} du
  Foresto}, {Kellerer}, {ten Brummelaar}, {Reess}, \&
  {Ziegler}}]{Merand_2006_07_2}
{M{\'e}rand}, A., {Coud{\'e} du Foresto}, V., {Kellerer}, A., {et~al.}
  2006{\natexlab{a}}, in SPIE Conference Series, Vol. 6268, 46

\bibitem[{{M{\'e}rand} {et~al.}(2005{\natexlab{b}}){M{\'e}rand}, {Kervella},
  {Coud{\'e} du Foresto}, {Ridgway}, {Aufdenberg}, {ten Brummelaar}, {Berger},
  {Sturmann}, {Sturmann}, {Turner}, \& {McAlister}}]{Merand_2005_07_0}
{M{\'e}rand}, A., {Kervella}, P., {Coud{\'e} du Foresto}, V., {et~al.}
  2005{\natexlab{b}}, \aap, 438, L9

\bibitem[{{M{\'e}rand} {et~al.}(2006{\natexlab{b}}){M{\'e}rand}, {Kervella},
  {Coud{\'e} du Foresto}, {ten Brummelaar}, \& {McAlister}}]{Merand_2006__0}
{M{\'e}rand}, A., {Kervella}, P., {Coud{\'e} du Foresto}, V., {ten Brummelaar},
  T., \& {McAlister}, H. 2006{\natexlab{b}}, MSAI, 77, 231

\bibitem[{{Moskalik} \& {Gorynya}(2005)}]{Moskalik_2005_06_0}
{Moskalik}, P. \& {Gorynya}, N.~A. 2005, Acta Astronomica, 55, 247

\bibitem[{{Nardetto} {et~al.}(2004){Nardetto}, {Fokin}, {Mourard}, {Mathias},
  {Kervella}, \& {Bersier}}]{Nardetto_2004_12_0}
{Nardetto}, N., {Fokin}, A., {Mourard}, D., {et~al.} 2004, \aap, 428, 131

\bibitem[{{Nardetto} {et~al.}(2009){Nardetto}, {Gieren}, {Kervella}, {Fouque},
  {Storm}, {Pietrzynski}, {Mourard}, \& {Queloz}}]{Nardetto_2009_05_0}
{Nardetto}, N., {Gieren}, W., {Kervella}, P., {et~al.} 2009, ArXiv e-prints

\bibitem[{{Neilson} {et~al.}(2010){Neilson}, {Ngeow}, {Kanbur}, \&
  {Lester}}]{Neilson_2010_06_0}
{Neilson}, H.~R., {Ngeow}, C., {Kanbur}, S.~M., \& {Lester}, J.~B. 2010, \apj,
  716, 1136

\bibitem[{{Perrin}(2003)}]{Perrin_2003_03_0}
{Perrin}, G. 2003, \aap, 400, 1173

\bibitem[{{Sachkov}(1997)}]{Sachkov_1997_10_0}
{Sachkov}, M.~E. 1997, Information Bulletin on Variable Stars, 4522, 1

\bibitem[{{Samus} {et~al.}(2009){Samus}, {Durlevich}, \& {et
  al.}}]{Samus_2009_01_0}
{Samus}, N.~N., {Durlevich}, O.~V., \& {et al.} 2009, VizieR Online Data
  Catalog: B/gcvs. Originally published in: Institute of Astronomy of Russian
  Academy of Sciences and Sternberg, State Astronomical Institute of the Moscow
  State University, 1, 2025

\bibitem[{{Storm} {et~al.}(2011){Storm}, {Gieren}, {Fouqu{\'e}}, {Barnes},
  {Pietrzy{\'n}ski}, {Nardetto}, {Weber}, {Granzer}, \&
  {Strassmeier}}]{Storm_2011_10_0}
{Storm}, J., {Gieren}, W., {Fouqu{\'e}}, P., {et~al.} 2011, \aap, 534, A94

\bibitem[{{ten Brummelaar} {et~al.}(2005){ten Brummelaar}, {McAlister},
  {Ridgway}, {Bagnuolo}, {Turner}, {Sturmann}, {Sturmann}, {Berger}, {Ogden},
  {Cadman}, {Hartkopf}, {Hopper}, \& {Shure}}]{ten-Brummelaar_2005_07_0}
{ten Brummelaar}, T.~A., {McAlister}, H.~A., {Ridgway}, S.~T., {et~al.} 2005,
  \apj, 628, 453

\end{thebibliography}

\end{document}